\documentclass[aip,apl,reprint]{revtex4-1} 
\usepackage{amsmath}
\usepackage{amssymb}
\usepackage{bm}
\usepackage{color}
\usepackage{graphicx}
\usepackage[latin1]{inputenc}
\usepackage{times}
\usepackage{natbib}
\makeatletter

\begin{document}


\title{Spin-torque oscillator linewidth narrowing under current modulation} 




\author{Ye.~Pogoryelov}
\email{yevgenp@kth.se}
\affiliation{Material Physics, Royal Institute of Technology, Electrum 229, 164 40 Kista, Sweden}

\author{P.~K.~Muduli}
\affiliation{Physics Department, University of Gothenburg, 412 96 Gothenburg, Sweden}

\author{S.~Bonetti}
\affiliation{Material Physics, Royal Institute of Technology, Electrum 229, 164 40 Kista, Sweden}

\author{Fred~Mancoff}
\affiliation{Everspin Technologies, Inc., 1300 N. Alma School Road, Chandler, AZ 85224, USA}

\author{Johan~\r{A}kerman}
\affiliation{Material Physics, Royal Institute of Technology, Electrum 229, 164 40 Kista, Sweden}

\affiliation{Physics Department, University of Gothenburg, 412 96 Gothenburg, Sweden}


\date{\today}

\begin{abstract}

We study the behavior of the linewidth of a nano-contact based spin torque oscillator (STO) under application of a radio frequency (100 MHz) modulating current. We achieve a significant (up to 85\%) reduction of the STO linewidth when it is modulated across a region of high nonlinearity. The mechanism responsible for the linewidth reduction is the nonlinear frequency shift under the influence of current modulation, which reduces the nonlinear amplification of the linewidth. The reduction of the linewidth during modulation can be quantitatively calculated from the free-running behavior of the STO. 

\end{abstract}


\maketitle 


Spin torque oscillators (STOs)\cite{Silva2008JMMM,SlavinTutorial} currently receive a rapidly growing interest due to their attractive combination of a large frequency range,~\cite{bonetti2009apl} high modulation rate,~\cite{Pufall2005APL,muduli2010prb} good scalability,~\cite{Villard2010IEEE} and fabrication processes similar to Magnetic Random Access Memory.~\cite{EngelBN2005,Akerman2005} The operating principle of the STO is the transfer of angular momentum from spin polarized electrons to the local magnetization of a thin magnetic layer, leading to large-amplitude coherent spin wave emission, which generates a microwave voltage signal through the magnetoresistance of the device.~\cite{Slonczewski1996,Berger1996} STOs derive their large frequency tunability from a very strong frequency-power nonlinearity,~\cite{slavinIEEE2005} which couples amplitude and phase fluctuations and leads to a spectral linewidth ($\Delta f$) that is significantly larger than the thermal broadening. The large tuning range hence comes at a cost of an amplified $\Delta f$, which has been intensely studied, both theoretically~\cite{Kim2006prb,Tiberkevich2008PRB,Kim2008PRL100a,Kim2008PRL100p,Carpentieri2010,Silva2010ieee} and experimentally,~\cite{Sankey2005PRB,Mistral2006,Boone2009PRB,Georges2009prb, Schneider2009prb,Bianchini2010a,Quinsat2010a} and must be further reduced for actual use of STOs in future communication applications. Another requirement is that STO signals need to be effectively modulated, which has been experimentally demonstrated both in single STOs~\cite{Pufall2005APL,muduli2010prb} and in synchronized STO pairs,~\cite{pogoryelov2010} and is well described by a combined nonlinear frequency and amplitude modulation (NFAM) theory.~\cite{consolo2010ieee,muduli2010prb} A direct consequence of nonlinear modulation is a shift of the carrier frequency, which affects the intrinsic STO nonlinearity. Since the nonlinearity governs the amplification of the thermal broadening, this opens up for the intriguing possibility of using modulation as an \emph{extrinsic} tool for controlling $\Delta f$. In this work we predict, and experimentally confirm, that modulation of a STO can indeed lead to a dramatic linewidth reduction (up to 85\%), and that the strong frequency-power nonlinearity, typically considered an intrinsic STO property, can be controlled by external means.

The device presented here consists of a 130 nm diameter nano-contact, fabricated on top of a Co$_{81}$Fe$_{19}$(20 nm)/Cu(6 nm)/Ni$_{80}$Fe$_{20}$(4.5 nm) pseudo spin valve stack. 
See Ref.~\onlinecite{Mancoff2006} for device fabrication details. All measurements were performed at room temperature with an external magnetic field of $\mu_{0}H=1$~T applied at an angle 75$^{\circ}$ to the film plane in order to only excite the propagating spin wave mode.~\cite{Slonczewski1999,Slavin2005PRL,Bonetti2010prl}

\begin{figure}[t!]
\centering
\includegraphics[width=0.45\textwidth]{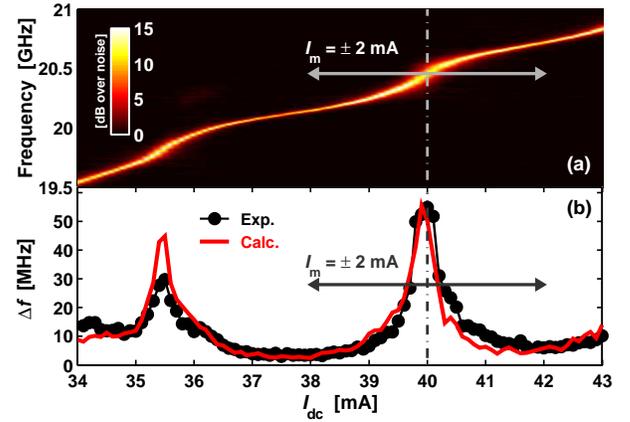}
\caption{(Color online) (a) Color map showing the current dependent STO frequency and power; (b) the corresponding linewidth. The solid line is a linewidth calculation based on ~\eqref{Eq1} using the current dependence of the STO frequency in (a).}
\label{fig:fig1}
\end{figure}

In Fig.~\ref{fig:fig1}(a) we show the free-running current dependence of the STO frequency with two regions of particularly strong frequency nonlinearity. Such large nonlinearities are common in nano-contact STOs and while their origin is still somewhat unclear, they are likely related to transitions between sub-modes of the main propagating spin wave mode.~\cite{Rippard2006PRB} Associated with both non-linear regions is a peak in $\Delta f$, shown in Fig.~\ref{fig:fig1}(b), which at $I_{dc}=40$~mA is over an order of magnitude larger than the minimum value around $I_{dc}=38$~mA. Assuming that the nonlinear damping term $Q=0$,~\cite{Georges2009prb, Boone2009PRB} $\Delta f$ can be expressed as~\cite{Kim2008PRL100a}
\begin{equation}
\Delta f = \Gamma _{G} \frac{k_{B}T}{E_{0}}\left ( 1+ \nu^{2} \right )
\label{Eq1}
\end{equation}

\noindent where $\Gamma _{G}$ represents the half-linewidth of the linear ferromagnetic resonance (FMR), $k_{B}$ is the Boltzmann constant, $T$ is the absolute temperature, $E_{0}$ is the average energy of the STO precession, $\nu = \left(I_{dc}/\Gamma _{G}\right)\left(df/dI_{dc}\right)$ is the STO nonlinearity, and $df/dI_{dc}$ is the local frequency tunability \emph{w.r.t.} current.

In Fig.~\ref{fig:fig1}(b) we show the calculated $\Delta f$ (red line) based on Eq.~\eqref{Eq1} using the experimental current dependence of the STO frequency [Fig.~\ref{fig:fig1}(a)], where we have estimated the value of $\Gamma _{G}\approx0.75$~GHz from a linear extrapolation of $\Delta f$ to zero current (not shown).~\cite{Georges2009prb} $E_{0}\approx1.7\cdot 10^{-17}$~J is then found from a fit of Eq.~\eqref{Eq1} to the experimental data. Knowing $E_{0}$, it is now possible to estimate the effective volume $V_{eff}$ of the magnetic material involved in the precession, using the expression:~\cite{Kim2008PRL100p}
\begin{equation}
E_{0}=\frac{V_{eff}M_{0}}{\gamma} f_{0}(\zeta -1)/(\zeta+Q)
\label{Eq2}
\end{equation}

\noindent where $\mu_{0}M_{0}=0.85$~T is the magnetization of the free layer, $\gamma$ is the gyromagnetic ratio, $f_{0}$ is the FMR frequency, $\zeta=\left(I_{dc}/I_{th}\right)$ is the supercriticality parameter with $I_{th}=23$~mA being the threshold current determined from a linear fit of the inverse power below threshold,\cite{Tiberkevich2007apl} and $Q$ is again set to zero. For our experimental geometry $f_{0}$ is estimated to be 15.5~GHz based on Eq.~(104a) of Ref.~\onlinecite{SlavinTutorial}. Finally, $V_{eff}$ is found to be $\approx120\cdot 10^{3}$~nm$^{3}$, which is about 2 times larger than the nominal volume of magnetic material under the nano-contact.

\begin{figure}[t]
\centering
\includegraphics[width=0.45\textwidth]{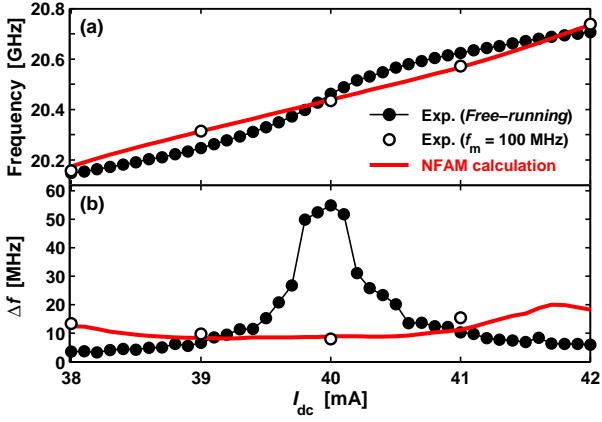}
\caption{(Color online) (a) Current dependence of the STO frequency for $I_{m}=0$~mA (filled circles) and $I_{m}=2$~mA (hollow circles), together with a NFAM calculation for $I_{m}=2$~mA (solid line); (b) Corresponding $\Delta f$ for current dependencies in (a), where the solid line is based on Eq.~\eqref{Eq1} and the $df/dI_{dc}$ calculated from (a).}
\label{fig:fig2}
\end{figure}

The frequency modulation experiment was performed by injecting a signal with frequency $f_{m}=100$~MHz and amplitude up to $I_{m}=2$~mA from a radio frequency source to the STO through a broadband microwave circulator. For more details see Ref.~\onlinecite{muduli2010prb}. Experimental set-up schematics is similar to the one depicted in Fig.~3.5 in Ref.~\onlinecite{Bonetti2010a}. The horizontal arrow in Fig.~\ref{fig:fig1}(a,b) shows the region across which the STO is modulated.

We first focus on the region around $I_{dc}=40$~mA, which has the greatest nonlinearity, and plot $f$ vs. $I_{dc}$ in Fig.~\ref{fig:fig2}(a), both in the free-running case and under modulation at $I_{m}=2$~mA. Using the free-running data, we then calculate the expected modulated response by using Eq.~(4) in Ref.~\onlinecite{muduli2010prb} and find an excellent agreement between our calculation (solid line) and the experimental data under modulation.

Fig.~\ref{fig:fig2}(b) shows the corresponding STO linewidth, both in the free-running case and under modulation. In addition, we can now make use of the calculated $f$ vs.\ $I_{dc}$ curve for the modulated STO, and utilize Eq.~\eqref{Eq1} to estimate $\Delta f$ under modulation. The calculation again agrees very well with the experimental data and both show a dramatic reduction of $\Delta f$ from well above 50 MHz to about 8 MHz, i.e. by about 85\%.

\begin{figure}[t]
\centering
\includegraphics[width=0.45\textwidth]{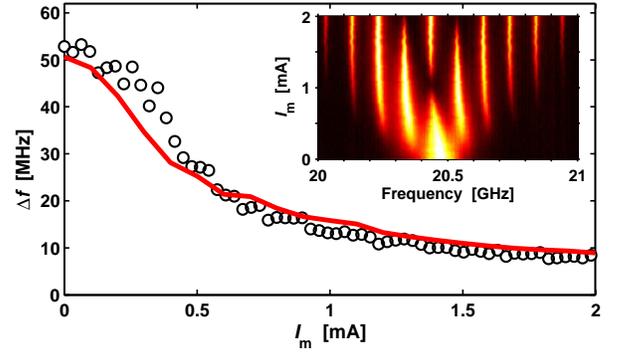}
\caption{(Color online) Measured (hollow circles) and calculated (solid line) $\Delta f$ as a function of the modulating current at $I_{dc}=40$~mA. Inset shows the development of sidebands with the modulation current.}
\label{fig:fig3}
\end{figure}

In Fig.~\ref{fig:fig3}, we show how the reduction in $\Delta f$ depends on the modulation current. The inset in Fig.~\ref{fig:fig3} shows the corresponding spectra as a function of the modulating current amplitude. 
Our results thus suggest that the observed $\Delta f$ at any modulation condition is almost entirely determined by the non-linear current dependence of the frequency, through the amplification given by Eq.~\eqref{Eq1}, and can be controlled by reducing this non-linearity through modulation, as predicted by the NFAM theory.

We hence conclude that $\Delta f$ can be externally reduced by the linearizing, or averaging, effect that the modulation has on the original strong nonlinearity. Similar agreement was found at $f_{m}=50$~MHz, which is consistent with the independence of the frequency shift on the modulation frequency (see Eq.~(4) in Ref.~\onlinecite{muduli2010prb}), and further corroborates that the mechanism is indeed the frequency shift brought about by the modulation.

\begin{figure}[t!]
\centering
\includegraphics[width=0.45\textwidth]{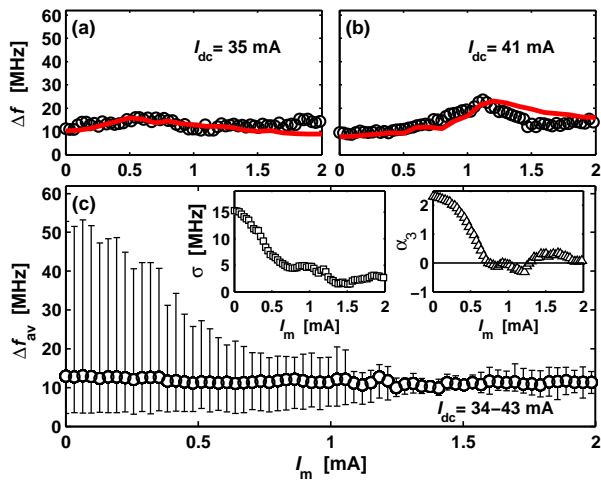}
\caption{(Color online) Linewidth as a function of the modulating current at two operating points (a) $I_{dc}=35$~mA and (b) $I_{dc}=41$~mA, together with the calculated $\Delta f$ (solid line); (c) average $\Delta f$ (hollow circles) and $\Delta f$ range (error bars) taken over the whole operating region, for increasing modulation current. Left inset: standard deviation of $\Delta f$. Right inset: skewness ($\alpha_{3}$) of the $\Delta f$ distribution.}
\label{fig:fig4}
\end{figure}

In the analysis above, we deliberately considered a region of particularly high nonlinearity, to show the strongest reduction in $\Delta f$ as a function of modulation. At other operating points, the modulating current may however result in much smaller reduction, no reduction, or even a slight \emph{increase} of $\Delta f$. For example, when modulated around $I_{dc}=35$~mA no significant change in the $\Delta f$ is observed over the whole range of modulating currents [Fig.~\ref{fig:fig4}(a)]. On the other hand, modulation around $I_{dc}=41$~mA reveals a local \emph{maximum} of the $\Delta f$ at $I_{m}=1$~mA. These observations are however still in good agreement with our previous analysis, and the observed behavior of $\Delta f$ can again be quite accurately predicted by NFAM theory and Eq.~\eqref{Eq1}, as shown by the calculated solid lines in Fig.~\ref{fig:fig4}(a,b). The calculations not only predict the correct trend of $\Delta f$ under modulation at each operating point, but also produce reasonable quantitative agreement with the experiment.

While it is hence true that modulation does not necessarily lead to a reduced $\Delta f$ at all operating conditions, it will always result in a global reduction of nonlinearities over the entire current range. For example, $\Delta f$ under modulation will never increase above the average value of the free-running $\Delta f$ over the modulated range ($\pm2$~mA). At the same time the average $\Delta f$ over the whole range of \textit{dc} current ($34-43$~mA) remains nearly constant under modulation at $f_{m}=100$~MHz [Fig.~\ref{fig:fig4}(c)]. This result suggests that on the average modulation will \emph{always} lead to narrowing (or no change) of $\Delta f$, since modulation has the same effect as local averaging, and any averaging over a data set can never have a larger standard deviation than the original data set. This is supported by the plot of the standard deviation of the linewidth over the whole range of \textit{dc} currents under study as a function of modulating current (shown as error-bars and in left inset in Fig.~\ref{fig:fig4}(c)). Right inset in Fig.~\ref{fig:fig4}(c) shows the skewness of the $\Delta f$ distribution as a function of the modulating current. High level of skewness at low modulating currents comes from the large nonlinearities related to mode transitions. Reduction of both the standard deviation by almost 85\% compared to the un-modulated state, and the skewness to almost zero at higher ($I_{m}>0.5$~mA) modulation currents means that modulation has the strongest impact on the initial large nonlinearities. These results confirm the increase of the linearity of the current dependent frequency behavior of a STO under modulation.

In conclusion, we have studied the behavior of the STO linewidth under modulation. We find that modulation has a general averaging effect on the nonlinearity of the STO signal, which is very well described by the theory for nonlinear modulation. Since the frequency nonlinearity is responsible for the local amplification of the linewidth, the reduction of this nonlinearity also leads to a dramatic reduction in the observed linewidth, which at some operating conditions was found to be as large as 85\%. The observed linewidth narrowing under modulation hence confirms both the NFAM theory and the auto-oscillator theory for linewidth amplification in STOs. 
Modulation can hence be used to both increase the linearity of a STO, suppress local variations of its signal properties to that of its global mean, and reduce the detrimental impact of mode transitions, all of which will be important for possible future communication and signal-processing applications of STOs.

Support from the Swedish Foundation for Strategic Research (SSF), the Swedish Research Council (VR), the G\"{o}ran Gustafsson Foundation and the Knut and Alice Wallenberg Foundation is gratefully acknowledged. Johan~\r{A}kerman is a Royal Swedish Academy of Sciences Research Fellow supported by a grant from the Knut and Alice Wallenberg Foundation.


%


%

\end{document}